\documentclass[english,twocolumn,pre]{revtex4-1}
\usepackage[latin9]{inputenc}
\usepackage{babel}
\usepackage{amsmath}
\usepackage{amssymb}
\usepackage{graphicx}
\usepackage{float}
\usepackage{titlesec}
\usepackage{tikz}
\usetikzlibrary{shapes,arrows}
\usepackage{lipsum}

\usepackage[export]{adjustbox}

\makeatletter

\usepackage{babel}
\begin{document}

\title {Simulating defect textures on coalescing nematic shells }
\date{\today}

\author{Badel L. Mbanga}
\email{badel.mbanga@tufts.edu}

\author{Kate K. Voorhes}

\author{Timothy J. Atherton}
\email{timothy.atherton@tufts.edu}
\affiliation{Department of Physics and Astronomy, Center for Nanoscopic Physics, Tufts University, 4 Colby st, Medford, MA 02155}

\selectlanguage{english}%


{\normalfont\large\bfseries}{\thesubsection}{}{}

\begin{abstract}
Two nematic shells brought in contact coalesce in order to reduce their interfacial tension. This process proceeds through the creation of a liquid neck-like bridge through which the encapsulated fluid flows. Following this topological transition,  We study the defect textures  as the combined shell shape evolves. Varying the sizes of the shells, we perform a quasistatic investigation of the director field and the defect valence on the doublet. Regimes are found where positive and negative defects exist due to the large negative Gaussian curvature at the neck. Using large scale computer simulations, we determine how annihilating defect pairs on coalescing shells are selected, and the stage of coalescence at  which annihilation occurs.\\\end{abstract}

\maketitle

\section{Introduction}


\footnotetext{\textit{~Department of Physics and Astronomy, Center for Nanoscopic Physics, Tufts University, 4 Colby st, Medford, MA 02155. Fax: +1 617-627-XXXX; Tel: +1 617-627-XXXX}}
\footnotetext{\textit{$^{\ast}$~E-mail: badel.mbanga@tufts.edu }}
\footnotetext{\textit{$^{\ddag}$~E-mail: timothy.atherton@tufts.edu }}

Studies of orientational order on curved manifolds have been revisited a great deal in the last decade \cite{nelson_colloids,vitelli_turner,vitelli_turner2, santangelo,vitelli_nelson,shin, nieves, nelson_peliti,prost_lubensky}. This is partly due to the realization that our current experimental capabilities make it possible to exploit the coupling of topological defects to curvature for a multitude of applications ranging from drug delivery systems to structuring in food and cosmetic products \cite{nelson_colloids}. The availability of massive computer simulations techniques has also played a major role in advancing our understanding of two-dimensional order confined to complex geometries \cite{shin, selinger, mbanga, subas}. 

While there is a great deal of literature on how to predict the ground state of the defect texture on an arbitrary, frozen,  geometry \cite{nelson_peliti,prost_lubensky,topcolloids}, a significant challenge is our ability to describe the routes to annihilation, coalescence or simply relocation  of topological defects as the host interface is deformed, though efforts in that vein are promising \cite{Lopez_trajectories, Bates,Lopez_nature}. A physically experimentally relevant system for such investigations is the assembly formed during the coalescence of two nematic shells. A single nematic shell is known to have a net topological charge of $+2$ \cite{Poincare}, with the four defects of charge $+ 1/2$ preferentially arranged in a tetrahedral configuration \cite{prost_lubensky}, or along great circles \cite{shin}, depending upon the relative strengths of the splay and bend elastic constants. Bringing two such shells in contact causes a rupture of the interface and the formation of a neck-like structure, thus facilitating the flow of the fluid contained within them. As these shells usually exist in a host, e.g.\ aqueous environment \cite{nieves}, interfacial tension would be minimized and the resulting shell should assume a spherical shape as well. At all stages of this coalescence process (after the initial contact), the net topological charge of the assembly must remain $+2$ as it is homeomorphic to a sphere. Hence the additional $+1/2$ defects must be annihilated during the coalescence process by nucleated $-1/2$ defects. There is a well-established theory of how topological defects couple to Gaussian curvature \cite{vitelli_turner,vitelli_turner2, santangelo,vitelli_nelson}; from this, we understand that four additional, negative charges that balance four of the existing positive charges are nucleated at the neck which has a large, negative Gaussian curvature. However, we have no immediate knowledge of the mechanisms involved in selecting which pairs of positive and negative disclinations annihilate. In this paper, we describe the adiabatic trajectories of defects as the shells coalesce using Monte-Carlo simulated annealing. The paper is organized as follows: in section II, we briefly revisit some key properties of orientational order on curved surfaces. In section III, we  present the geometry of the shells studied numerically. Section IV describes the numerical techniques employed, and the resulting ground states are discussed in section V, followed by a brief conclusion.    

\section{  Orientational order on curved interfaces}
Curved interfaces with nematic-like order show a dependence of the ground state of the nematic texture on the geometry of the interface; moreover, the mechanical stability and the nematic texture are co-dependent \cite{selinger_soft_matter}. If the surface is parametrized by $X=X(u,v)$, and we consider the case where the director is confined to lie tangent to the surface of the shell, then we can define an orthonormal coordinate system $\{\hat{e}_u,\hat{e}_v,\hat{N}\}$ such that the local nematic director is defined as $\hat{n} = \cos \theta \hat{e}_u+ \sin \theta \hat{e}_v$. If the interface is a developable surface viz. a plane or a cylinder, the nematogens orient uniformly and parallel to one another, that is, the ground state is a defect-free configuration. However, if the interface has nonzero Gaussian curvature $G\neq 0$, one immediately sees that the defect-free texture is no longer attainable. To illustrate this, consider a director field $ \theta(u,v)$ confined to lie tangent to a given surface with non-vanishing Gaussian curvature. One can write the Frank-Oseen elastic distortion energy in the one-constant approximation as 

\begin{equation}
F = \dfrac{K}{2}  \int dA \; ( \partial_i \theta - A_i )( \partial_j \theta - A_j )
\end{equation}

Here, $K$ is the Frank elastic constant, and $ A$ is the spin connection that captures rotations of the tangent frame along the surface. It is defined such that $\nabla \times A = - G$.  A uniform director orientation is possible only if the integrand vanishes, \i.e.  $\partial_i \theta - A_i  = 0$ everywhere. Applying the 2D antisymmetric  Levi-Civita tensor to both sides of this expression, one finds $ \epsilon_{ij} \partial_i \theta - \epsilon_{ij} A_i  = 0$. The first term vanishes, yielding $\epsilon_{ij} A_i  = 0$, i.e.\ $G = 0$, which contradicts the aforementioned hypothesis of nonzero Gaussian curvature; thus one expects mismatches in the orientation of the nematogens. These mismatches are disclinations that give rise to the Schlieren textures which can be observed under polarizing microscopy. 

In describing the defect texture, one can separate the different contributions to the energetics of the defects into the defect-defect interaction and the defect-curvature interaction. This is captured by the expression: 

\begin{equation}
E = \dfrac{C}{2} \sum_{m,n}s_n V_{d-d} (u_n-u_m)s_m + C \sum_n V_{G}(u_n)s_n(1-\dfrac{s_n}{4 \pi})
\end{equation}

The first term is the long-range interaction among the disclinations. $S_i$ is the charge of the $i^{th}$ disclination and $V_{d-d}(u)$ is the interaction potential which is obtained readily by inverting the Laplace-Beltrami operator. $C$ is the renormalized elastic constant, proportional to the Frank elastic constant. This electrostatic-like interaction drives a strong repulsion of like-charged defects. Disclinations interact with Gaussian curvature via a ``Geometric Potential" \cite{vitelli_turner} as illustrated by the second term, with $V_G = - \log \Omega(u)/2$; here, we have defined the conformal factor $\Omega(u)$ such that the metric tensor is expressed as $g_{ij} = \Omega(u) \delta_{ij}$.  This geometric potential causes defects to be attracted to the regions with the maximum positive or minimum negative Gaussian curvature, for positive and negative disclinations respectively \cite{vitelli_turner}. This expression is independent of the coordinate $v$ due to azimuthal symmetry.  

The different roles played by the intrinsic and extrinsic curvatures of the interface in determinining the ground state of the defect texture become particularly interesting for these interfaces with neck-like regions. The extrinsic geometry promotes a uniform orientation of the nematogens in those regions where the changes in the orientation of local normal are greatest, i.e.\ the regions with the largest Gaussian curvature or the neck. These two requirements, namely placement of defects at the neck and uniform director configuration, are incompatible and lead to a frustration of the orientational order. This is of great interest, both at the fundamental level and with regard to applications. The coupling of orientational order to curvature thus plays an important role in the structure and function of several two-dimensional systems, such as crystals \cite{Irvine}, superfluid films \cite{vitelli_turner2}, nematic-coated particles \cite{nelson_colloids,shin} and smectic films \cite{santangelo}. Cellular mechanisms such as material exchange between vesicles rely on the formation of stable neck structures  \cite{McMahon}, which are regions with large negative gaussian curvature, i.e.\ where one expects a very strong prevalence of extrinsic coupling. Having reviewed these applications, we now turn our attention to the geometry of interest.

\section{ Geometry of coalescing shells}


A great deal of the literature on the hydrodynamics of both early and late stages of droplets coalescence exists for both analytical \cite{Hopper,stone}, numerical \cite{stone,drops_geometry} and experimental \cite{aarts_bonn} studies. The parametrization used in this paper is that obtained by Garabedian and Helble \cite{drops_geometry}. 
\begin{figure}
\begin{center}
\includegraphics[height=4.75cm, angle=0]{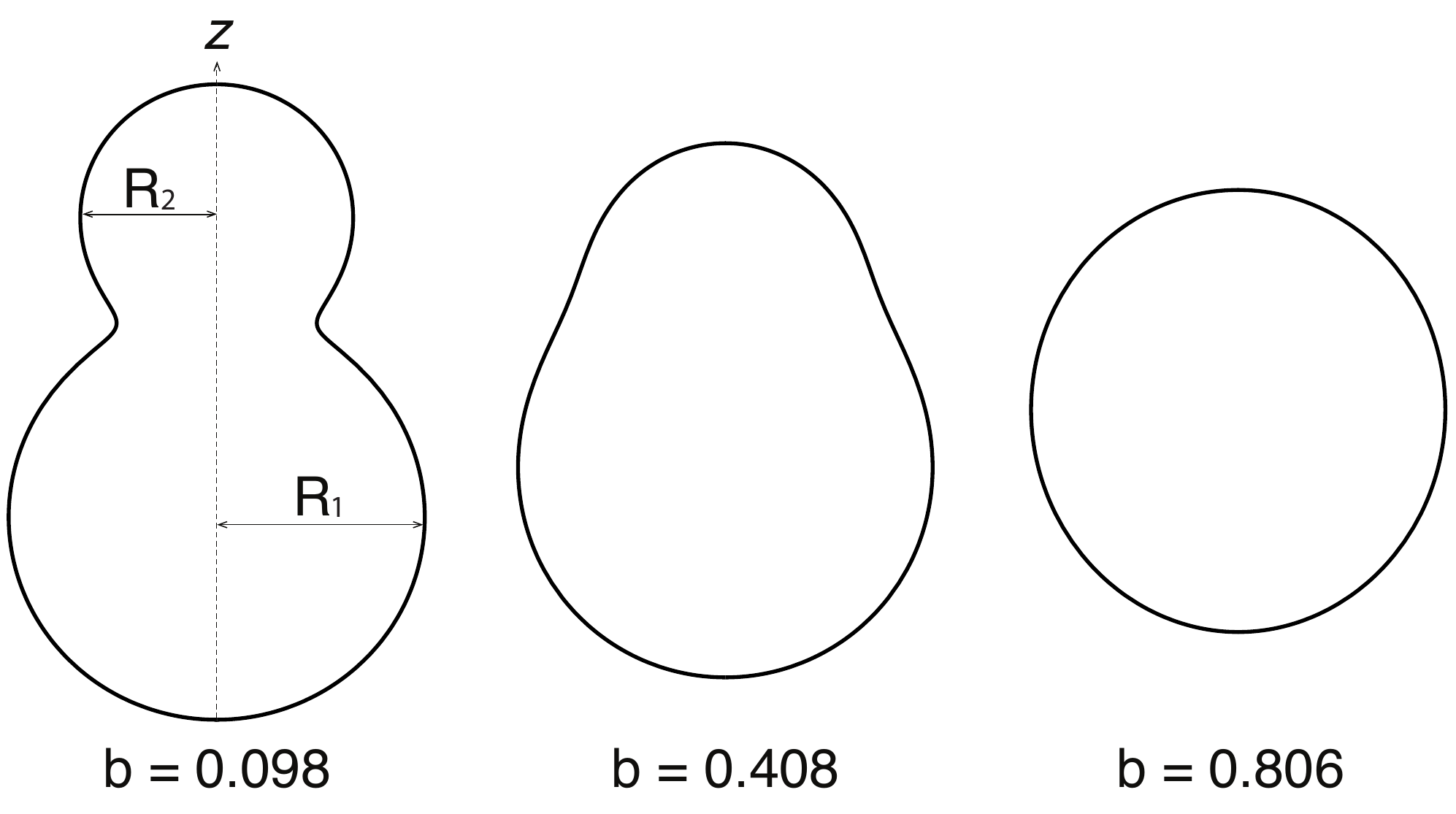} 
\caption{{Shape at different stages of coalescence of two shell of radii $R_1 = 1.5$ and $R_2 = 1$ (r=1.5). }}
\label{geometry}
\end{center}
\end{figure}  
For two coalescing shells of radii $R_1$ and $R_2$ and reduced radius $r=R_1/R_2$, there is a one-parameter family of surfaces  that spans from two shells in contact at one point to a single shell via a series of biconcave, pear-like, and ellipsoidal shapes. A normalized parameter $b$ controls the degree of coalescence where $b=0$ represents the two initial shells just beginning to touch, and $b=1$ the fully coalesced sphere. This parametrization has been shown to adequately capture the shape of coalescing droplets of fluids with viscosities spanning a wide range \cite{drops_geometry}. The dependence of the coalescence parameter $b$ on the coalscence time, the growth rate of the neck, and the total surface area of the doublet are discussed at length in  \cite{drops_geometry}. Fig. \ref{geometry} illustrates a typical shape evolution of the assembly formed by two coalescing shells. 

\section{ Monte Carlo simulations }
Throughout our discussion, we will make the simplified assumption that the rearrangement of the defects occurs at a timescale much faster than that of the elastic relaxation of the shell assembly. This warrants a quasi-static approach in which we will investigate the ground state of the defect texture in a sequence of frozen geometries obtained from the ansatz surface discussed in the previous section that reproduces qualitatively the shapes obtained in experiments on coalescing shells. This proves computationally more tractable --and allows us to separate geometric from dynamic effects-- than a full dynamic study. 
A very well established approach to computer simulations of particles with nematic order is the Lebwohl-Lasher model \cite{Lebwohl_Lasher} which is a simple an elegant discrete formulation of the Maier-Saupe theory of nematic liquid crystals. It simply favors parallel alignment of neigbouring particles. This approach captures the essential features of the nematic-isotropic phase transition in three dimension, and the KT transition in two dimension as well. We use a modified version of the model that accounts for the nonuniformity of the discretization, which is a inevitable consequence of the varying curvature \cite{mbanga}.

\begin{figure}
\begin{center}
\includegraphics[scale=.16, angle=0]{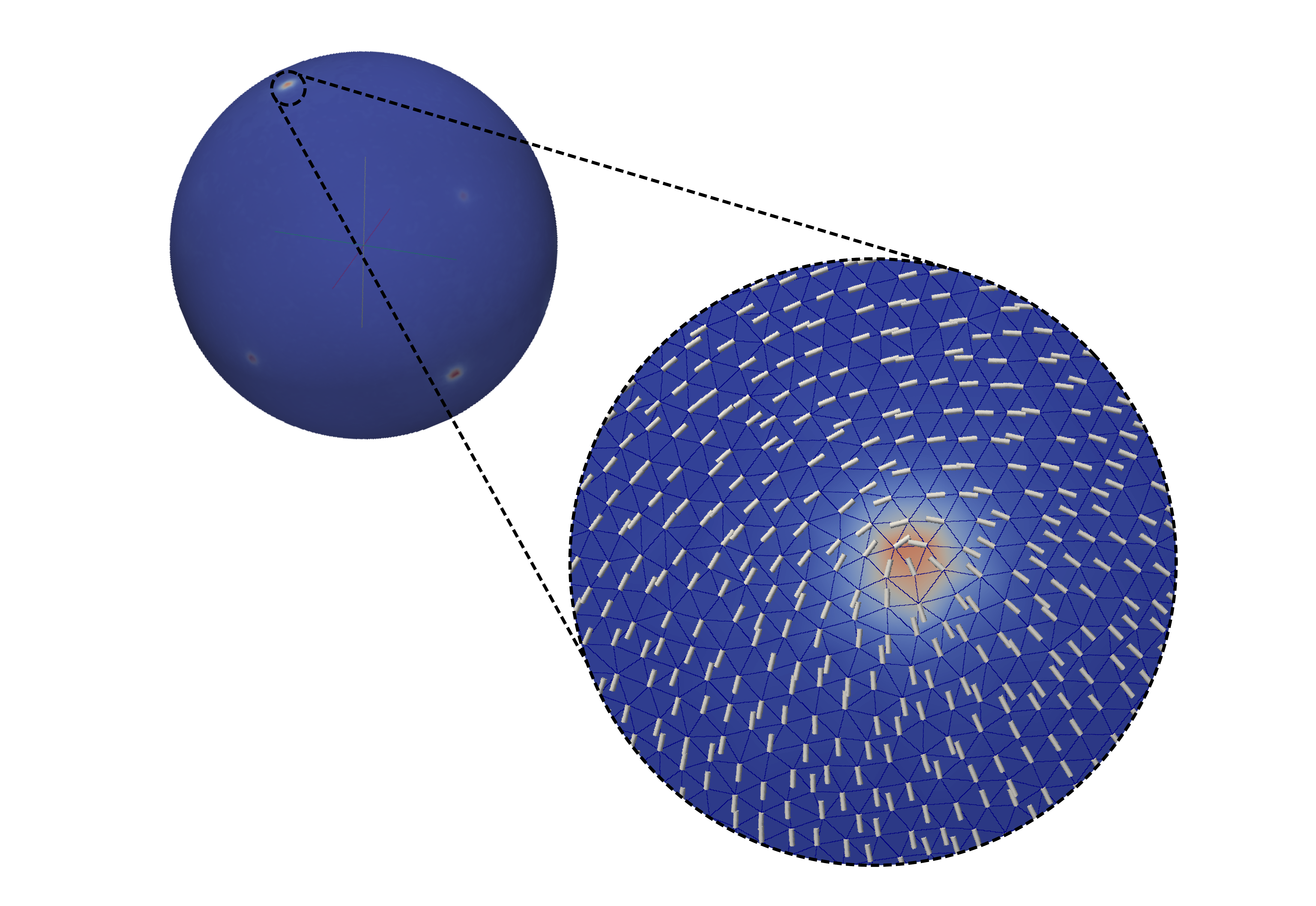} 
\caption{{Left: tetrahedral arrangement of the disclinations obtained from Monte-Carlo simulations. The discretization into triangular patches along with the director field near a defect is shown in the zoomed section on the right. }}
\label{defect_example}
\end{center}
\end{figure}

Starting with a surface discretized in a large number of small triangular patches,  a nematic director is assigned to the tangent plane of each patch.  In the one elastic constant approximation, the interaction energy is given by $H_{int} =  \sum_{i,j} \Delta A_i \Delta A_j \big[1 - ({\bf n}_i \cdot {\bf n}_j)^2 \big] V(r_{ij})$.
Here $V(r_{i j}) = \exp [-r_{i j}^2/(2 \sigma^2)]$ weights spin coupling between patches $i$ and $j$ by their spatial distance $r_{ij}$,  $\Delta A_i$ is the area of the $i^{th}$ patch and we choose $\sigma = 0.1 R_1$.  The area weighting has the important advantage that the coarse-grained model appropriately reduces to the continuum form of the energy in the limit that $\Delta A_i \to 0$, and minimizes the influence of nonuniform mesh geometry.   Fig. ~\ref{defect_example} shows the discretization on a patch of a sphere and typical result of the director configuration around a $s=1/2$ disclination. 

To obtain any given data point of our phase space $(r, b)$, we use a combination of a Monte-Carlo simulated annealing scheme with the Metropolis-Hastings sampling method and parallel tempering with $10$ replicas of the system; the configuration exchange allows the system to extract itself from local minima. We consistently find the final configuration obtained this way to be of lower energy than the best of a large number of independent simulations, and it has the added benefit that it does not require as many replicas as would ordinary parallel tempering. The lowest energy state obtained is accepted as the ground state for that parameter set. The final number of disclinations, their positions, and the director field minimizer are recorded for analysis.

\section{Defects valence and textures}

\begin{figure}[h]
\begin{center}
\includegraphics[height=5.75cm, angle=0]{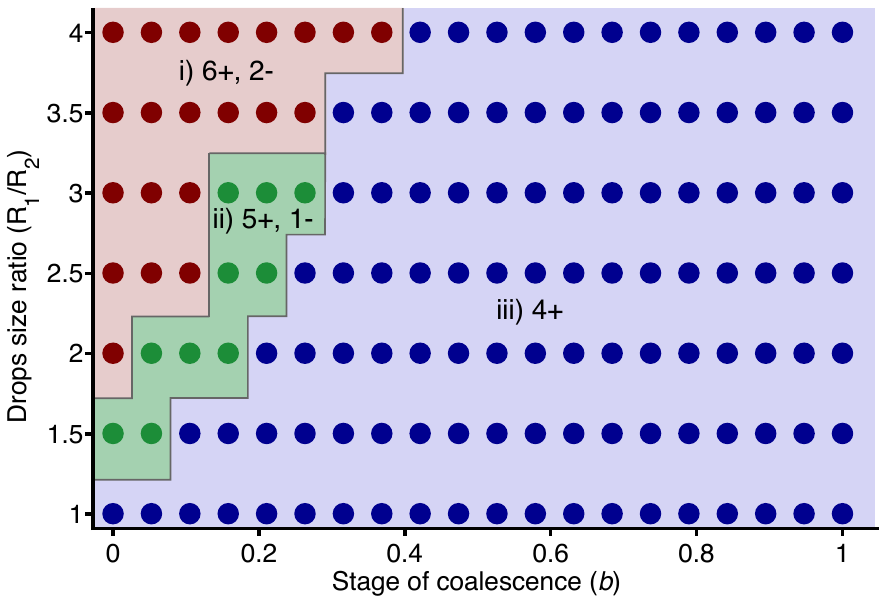} 
\caption{{ Defect valence at different stages of coalescence for shells of different aspect ratio. }}
\label{phase_diagram}
\end{center}
\end{figure}


\begin{figure*}
\centering
\includegraphics[height=12.0cm]{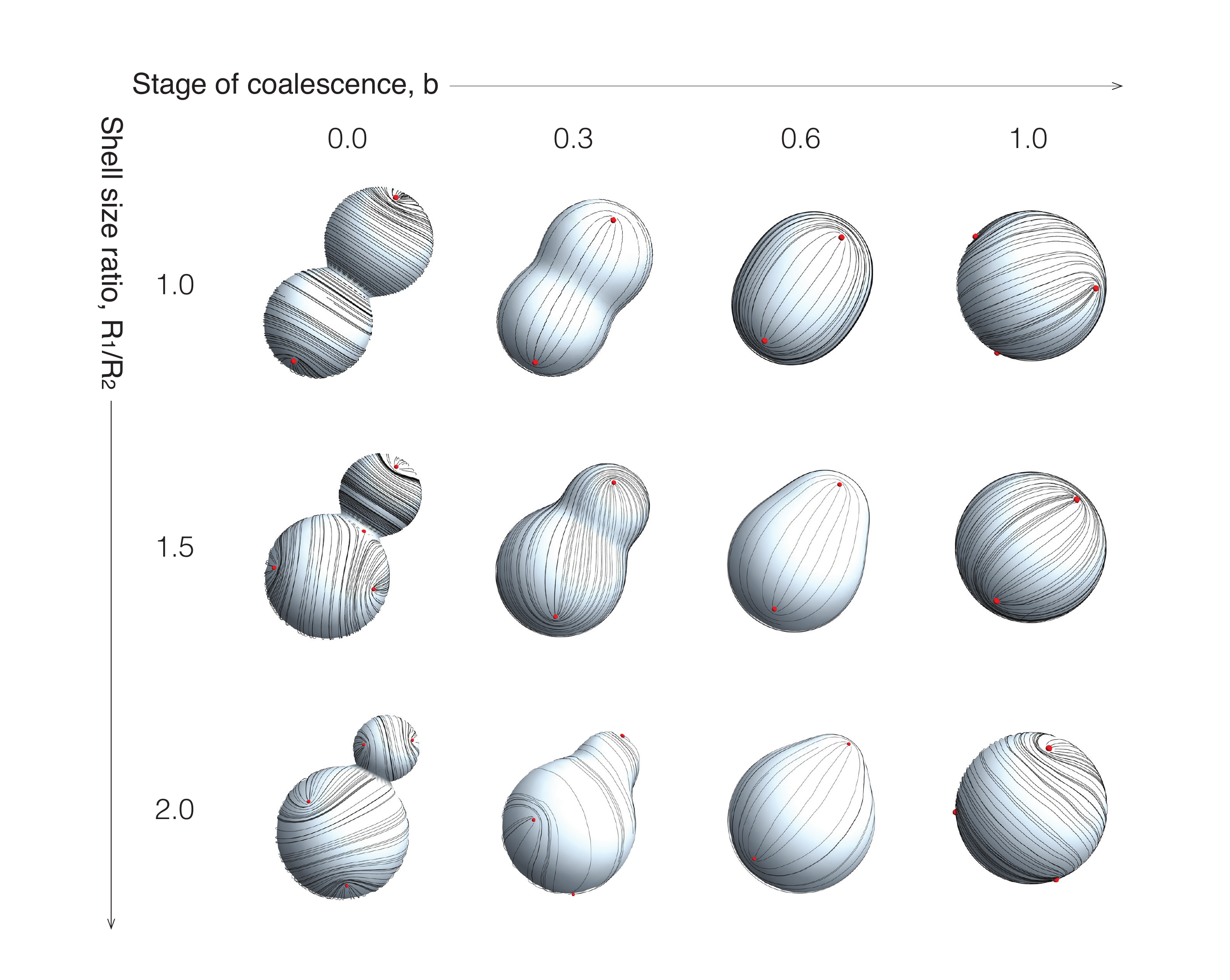} 
\caption{Defects arrangement on the doublet during coalescence: $r = R_1/R_2$ is the spheres relative size, $b$ is the stage of coalescence. For larger values of $r$, the defects arrangement becomes asymmetric with more disclinations located on the side with smaller Gaussian curvature. Red dots indicate the positions of the defects.}
\label{textures}
\end{figure*}

The requirement of a constant net topological charge ultimately leads to the annihilation of four positive defects with the four negative ones nucleated at the neck; however, at very early stages of coalescence, we find textures with coexisting positive and negative charges. This is due to the high concentration of negative Gaussian curvature at the neck and hence the geometric potential outweighs the attraction of the negative charges to the positive ones. We observe a sequence of transitions characterized by the number of positively charged disclinations, going from $6$, $5$, to $4$. The appropriate number of negative charges is present in each case, namely  $2$, $1$, and $0$ respectively. Textures with 8 or 7 positive defects might occur at the earliest stages, but were not observed in our simulations. The phase diagram in Fig. ~\ref{phase_diagram} shows the dependence of the number of defects on both the stage of coalescence and the shell size ratio. For equal sized shells, we only see four defects. It is immediate from Fig. ~\ref{phase_diagram} that annihilation occurs in the early stages of coalescence and that coexistence of opposite-charged defects becomes possible as more negative Gaussian curvature is confined in a region of the drops assembly, i.e.\ as the ratio of the drops size increases. A striking feature of this phase diagram is the apparent tricritical point around $3\leq r \leq3.5$ and $0.25 \leq b \leq 0.3$. It is worth noting that at a size ratio $r=3$ and coalescence stage $b \approx 0.25$, the texture shows two positive defects on the small shell, three positive defects on the large shell, and a single negative defect at the neck. As $b$ is increased to $b \approx 0.3$, a positive defect from the large droplet annihilates with the negative defect on the neck, resulting into a symmetric arrangement of the four remaining defects along a great circle, similar to the texture shown in Fig. ~\ref{textures} for $r=1$ and $b \approx 0.3$. On the other hand, at $r=3.5$ and $b \approx 0.25$, we find three positive defects on each of the two droplets, arranged into a triangular configuration; in this case, two negative defects are located at antipodal points on the neck. A small increment of $b$ to $b \approx 0.3$ results into an annihilation of two defects from the small droplet with those on the neck, and the final defect arrangement consists of a single defect on the small droplet and three defects arranged in a triangular configuration on the larger droplet.

As the coalescence proceeds, like-charged defects rearrange themselves on the surface to maximize their repulsion while satisfying their attraction to Gaussian curvature. Fig. ~\ref{textures} shows the director field and disclinations arrangement at different stages of coalescence for a number of drops assemblies.For $r=1$, the defects arrangement on the doublet remains symmetric with respect to the neck throughout coalescence; therefore, tracking the trajectory of defects simply amounts to finding the average distance of defects from the neck for each value of $b$ being simulated. On the other hand, the positioning of the defects for $r>1$ shows an asymmetry that is more pronounced the higher the value of $r$. The positive disclinations initially on the smaller shell are selected to annihilate with the negative ones at the neck. The ground state thus consists of, e.g.\ for $r \geq 3.5$, and $b \leq 0.3$, four positive disclinations on the lower curvature region, two positive disclinations on the higher curvature region, and two negative disclinations trapped in the neck region.



\begin{figure}
\includegraphics[height=6.0cm, angle=0]{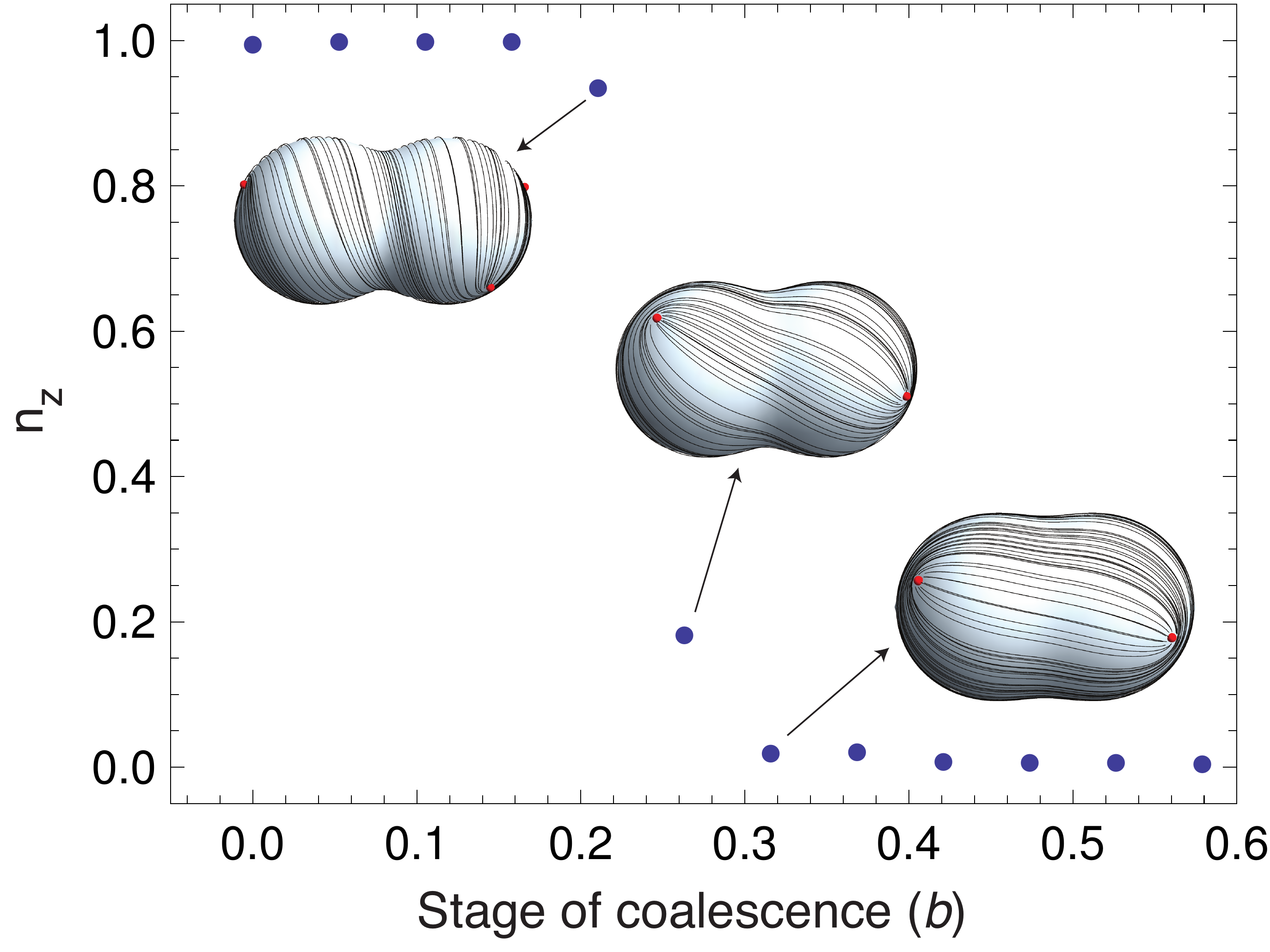} 
\caption{{Director configurations near the orientational transition from bend-rich to splay-rich; $r=1$, $b_c \approx 0.25$. $n_z$ is the component of the director along $\hat{e}_v$. The red dots show the positions of the defects.}}
\label{bend_splay_transition}
\end{figure}

The early stages of coalescence are characterized by a smooth orientational transition as shown in 
Fig. \ref{bend_splay_transition} for the case of equal-sized droplets. At the start of the coalescence process, the ground states obtained from our simulations display a preference for a bend-rich configuration with the director mainly oriented along the lines of latitude, or $\hat{e}_v$. A smooth orientational transition occurs starting at a critical value $b_c$, where the preferred director configuration starts switching from the bend-rich to a splay-rich configuration. In these early stages, whereas the regions far from the neck are still almost spherical, \i.e. the principal curvatures are equal, the neck shows a strong discrepancy in the scales of the principal curvatures, resulting in the director adopting the direction of lowest curvature, which itself changes with increasing $b$. A subsequent global reorientation of the director field to the well-known tetrahedral configuration occurs as the shape approaches a sphere.

\section {Conclusion}
Using Monte-Carlo simulated annealing, we have simulated the ground state texture of nematic shells undergoing coalescence. We have found ground states exhibiting a coexistence of positively and negatively charged disclinations when the geometric potential due to the Gaussian curvature concentrated at the neck dominates the attraction of oppositely charged disclinations. A rearrangement of the disclinations occurs during the coalescence, leading to the annihilation of extra disclination pairs, and an orientational transition; in the case of equal Frank elastic constants, as studied here, the tetrahedral arrangement of the disclinations is obtained upon completion of coalescence. A careful control of the defect valence and arrangement will enable construction of even more complex microstructures based on functionalization of defects spots than envisioned in \cite{nelson_colloids} for spherical colloids. Current microfluidics advances  \cite{Spicer} have already succeeded in arresting the coalescence of emulsion droplets through bulk jamming, thus stabilizing the non-spherical droplets or colloidal particles here studied. To isolate the role of geometry on the director textures In this study, a number of other physical effects were ignored. Firstly, the elastic constants were taken to be equal; we anticipate that for $K_{11} \neq K_{33}$, the position of the splay-rich to bend-rich transition will shift. Furthermore, we since we considered a quasi-static limit, it is likely that dynamic effects could substantially alter the annihilation process if the timescale of elastic relaxation becomes comparable to the timescale of coalescence. We are presently developing a model for such systems
\section* {Acknowledgements}
It is our pleasure to thank Chris Burke, Marco Caggioni and Patrick Spicer for fruitful discussions as we were performing this research. One of the authors (KKV) was supported by the undergraduate Summer Scholars program at Tufts University. Part of this work was completed at the University of New South Wales, Australia, during a residency funded by a Sandler International Research grant from Tufts University. Simulations were performed using Tufts University's HPC facilities.


\bibliography{nematicdrops_PRE}

\end{document}